\documentclass[aps,prl,groupedaddress,
showpacs,amsmath,amssymb]{revtex4}

\usepackage{graphicx}
\usepackage{bm}
\usepackage{bbm}

\def\[#1\]{\begin{align}#1\end{align}}
\def\sgn{\mathop{\rm sgn}}

\def\be{\begin{equation}}
\def\ee{\end{equation}}
\def\ba{\begin{eqnarray}}
\def\ea{\end{eqnarray}}
\def\bc{\begin{center}}
\def\ec{\end{center}}

\begin{document}

\title{Inter-valley plasmons in graphene}

\author{T. Tudorovskiy}
\altaffiliation[Also at]{
Fachbereich Physik der Philipps-Universit\"at Marburg, Renthof 5,
D-35032, Germany}
 \email{Timur.Tudorovskiy@physik.uni-marburg.de}
\author{S. A. Mikhailov}
 \affiliation{Institute of Physics,
University of Augsburg,
D-86135 Augsburg, Germany}

\date{\today}

\begin{abstract}
The spectrum of two-dimensional (2D) plasma waves in graphene has been recently studied
in the Dirac fermion model. We take into account the whole dispersion relation for graphene
electrons in the tight binding approximation and the local field effects in
the electrodynamic response. Near the wavevectors close to the corners of the hexagon-shaped Brillouin zone we
found new low-frequency 2D plasmon modes with a linear spectrum. These ``inter-valley'' plasmon modes are related to the transitions between the two nearest Dirac cones.
\end{abstract}

\maketitle

\begin{figure}
\begin{minipage}{80mm}
\begin{tabular}{c}
\includegraphics[height=5cm]{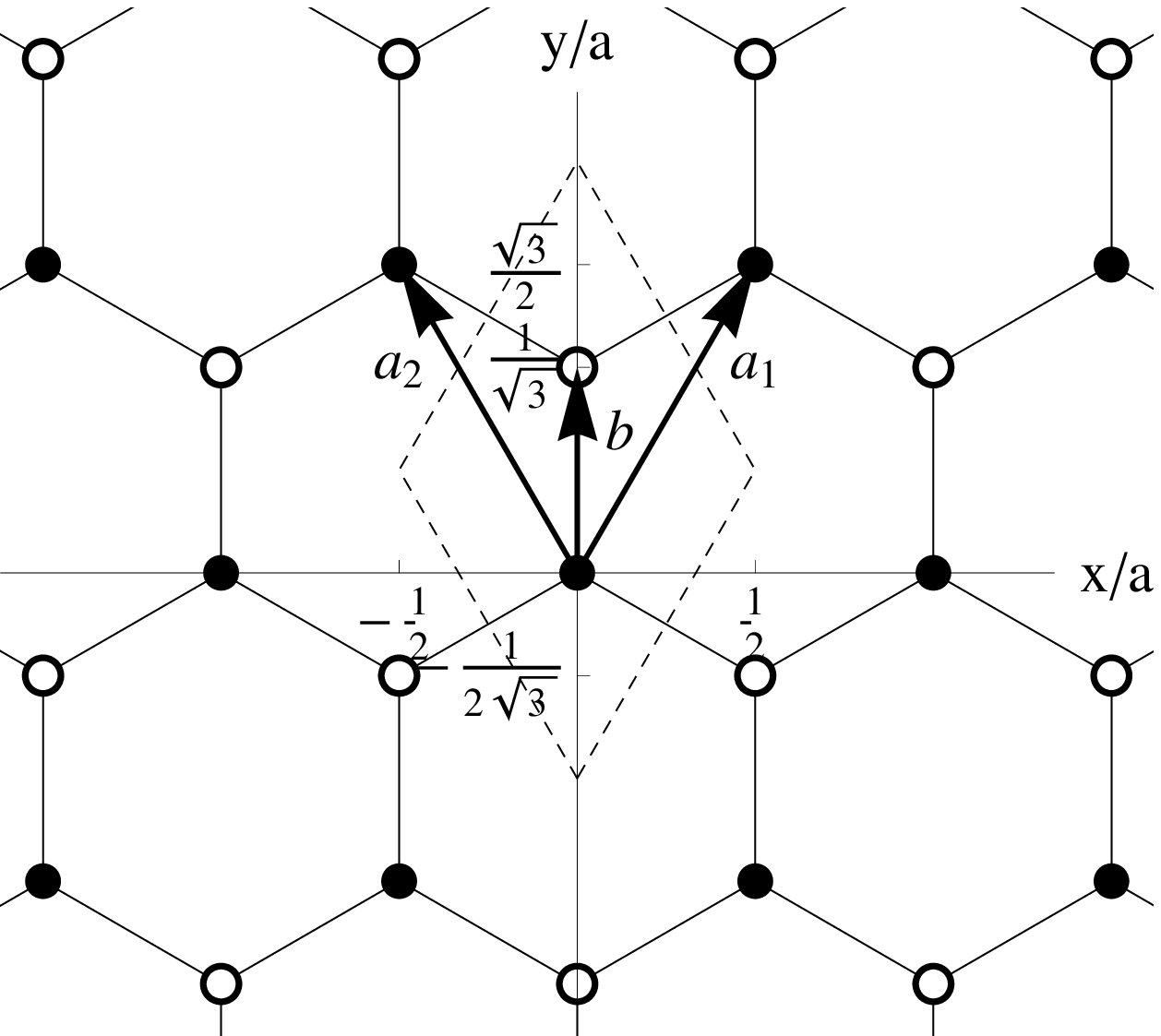}\\
(a)
\end{tabular}
\end{minipage}
\begin{minipage}{80mm}
\begin{tabular}{c}
\includegraphics[height=5cm]{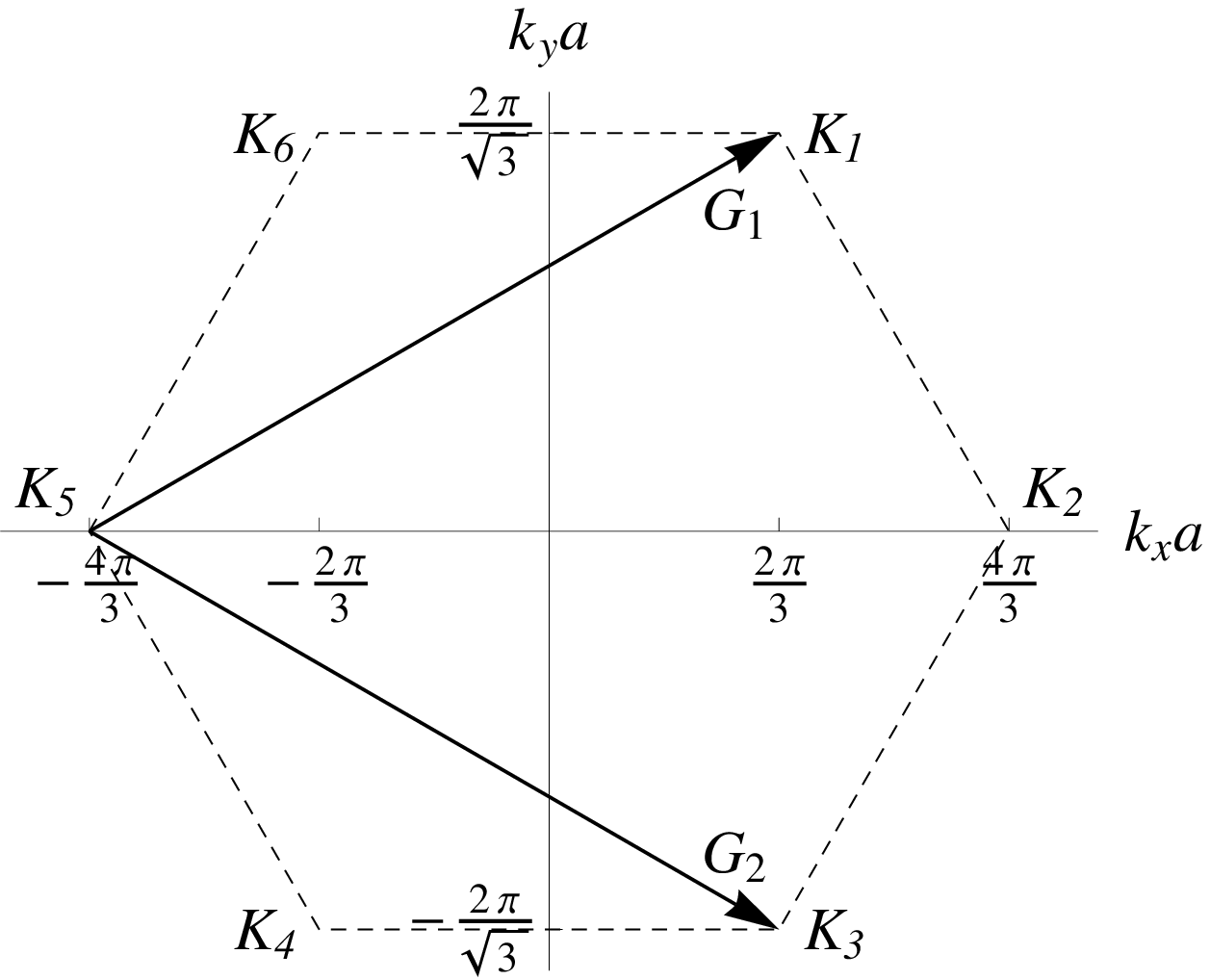}\\
(b)
\end{tabular}
\end{minipage}
\caption{\label{latt&BZ} (a) The honey-comb lattice of graphene. All points of the first sublattice (black circles) are given by $n_1{\bf a}_1+n_2{\bf a}_2$, of the second sublattice (open circles) by $n_1{\bf a}_1+n_2{\bf a}_2+{\bf b}$. (b) The BZ of graphene. The basis vectors of the reciprocal lattice are ${\bf G}_1$ and ${\bf G}_2$. ${\bf K}_i$, $i=1,\ldots,6$ are the corners of the BZ (the Dirac points). Dashed lines show the the boundaries of the elementary cell both in direct and reciprocal space. In the figure $\textbf{a}_1=a(1/2,\sqrt{3}/2)$, $\textbf{a}_2=a(-1/2,\sqrt{3}/2)$, $\textbf{b}=a(0,1/\sqrt{3})$, $\textbf{G}_1=2\pi a^{-1}(1,1/\sqrt{3})$, $\textbf{G}_2=2\pi a^{-1}(1,-1/\sqrt{3})$, $\textbf{K}_1=-\textbf{K}_4=2\pi a^{-1}(1/3,1/\sqrt{3})$, $\textbf{K}_2=-\textbf{K}_5=2\pi a^{-1}(2/3,0)$,
$\textbf{K}_3=-\textbf{K}_6=2\pi a^{-1}(1/3,-1/\sqrt{3})$, where $|\textbf{a}_1|=|\textbf{a}_2|=a$, $|\textbf{G}_1|=|\textbf{G}_2|=G=4\pi/(\sqrt{3}a)$.}
\end{figure}

Graphene, a recently discovered \cite{Novoselov05,Zhang05} two-dimensional (2D) material consisting of a single layer of carbon atoms, has been in the focus of experimental and theoretical research in the past years (see Ref.~\cite{cas09} and references therein). The carbon atoms in graphene form a dense 2D honeycomb lattice, Fig.~\ref{latt&BZ}a, with two atoms per elementary cell. The band structure of graphene electrons
\cite{wal47,Slonczewski58,sai98} consists of two bands
touching each other at six points ${\bf k}={\bf K}_i$, $i=1,\dots,6$ at the corners of the Brillouin zone (BZ, Fig.~\ref{latt&BZ}b).  In the vicinity of ${\bf K}_i$ the dispersion surface forms two cones with vertexes at ${\bf K}_i$. In the intrinsic graphene at zero temperature the lower (``hole'') band is fully occupied while the upper (``electron'') band is empty, and the Fermi level goes through the Dirac points.  Using the doping or applying a gate voltage between the graphene layer and a  substrate (in a typical experiment the graphene layer lies on a Si/SiO$_2$ substrate) one can shift the chemical potential $\mu$ to the electron or to the hole band and vary the density of electrons and/or holes.

Near the Dirac points the graphene quasi-particles have a linear, quasi-relativistic dispersion
\be
E_{{\bf k}l}^{Dir}=l\,\hbar V |{\bf k}-{\bf K}_i|, \quad |{\bf k}-{\bf K}_i|a\ll 1.
\label{spectrum}
\ee
Here $l=+1$ and $-1$ correspond to the electron and hole band respectively, $V\approx 10^8$ cm/s is the Fermi velocity in graphene and $a=2.46\textrm{\AA}$ is the lattice constant, Fig.~\ref{latt&BZ}a. It is the massless energy dispersion of graphene quasi-particles (\ref{spectrum}) that leads to its amazing physical properties and caused the great interest to this material.

In this Letter we address the problem of plasma oscillations in graphene. The plasma waves in graphene have been considered  in Refs. \cite{Hwang07,Wunsch06,Polini08,Vafek06,hil09}. In these publications the spectrum of plasma waves has been calculated in the long-wavelength limit $qa\ll 1$ from zeros of the Lindhard dielectric function \cite{Lindhard54},
\be
\epsilon_{Dir}({\bf q},\omega)=
1-\frac{2\pi g_v g_s e^2}{q\kappa S}\sum_{{\bf k}ll'}
\frac
{f(E^{Dir}_{{\bf k}l})-f(E^{Dir}_{\textbf{k}+\textbf{q},l'})}
{E^{Dir}_{{\bf k}l}-E^{Dir}_{\textbf{k}+\textbf{q},l'}+\hbar\omega+i0}
|\langle \textbf{k}+\textbf{q},l'| e^{i{\bf q r}} |{\bf k}l\rangle_{Dir}|^2,
\label{epsilon00}
\ee
which can be obtained within the self-consistent-field approach \cite{Ehrenreich59} or, equivalently, in the random phase approximation. Here $S$ is the area of the graphene sample, $f(E)$ is the Fermi-Dirac distribution function,
$-e$ is the electron charge ($e>0$),
$\textbf{q}$ is the wavevector of an electric field in the 2D plane, $q=|\textbf{q}|$, $g_s=g_v=2$ are the spin and valley degeneracies and $\kappa$ is the dielectric constant of surrounding medium. The wavefunctions $|{\bf k}l\rangle$ have been found from the Dirac approach, when the system is described (near the Dirac points) by the effective  Hamiltonian $H_{Dir}=V\sigma_\alpha \hat p_\alpha$, where $\alpha$ takes the values $x,y$, $\sigma_\alpha$ are Pauli matrixes, and $\hat p_\alpha$ is the momentum operator. The sub-/superscript ``Dir''  in (\ref{epsilon00}) reminds that the energies and the wavefunctions have been calculated within the Dirac (effective medium) approximation. In the limit $q\ll k_F$ the spectrum of 2D plasmons takes the form \cite{Hwang07,Wunsch06}
\be
\omega_p(q)=\left(\frac{e^2g_sg_v|\mu|}{2\hbar^2\kappa}q\right)^{1/2}=\left(\frac{e^2 V\sqrt{g_sg_v\pi n_s^{0}}}{\hbar\kappa}q\right)^{1/2},
\label{longwavepl}
\ee
which coincides with the standard 2D plasmon dispersion $\omega_p(q)=\sqrt{2\pi n_s^0 e^2 q/(m^\star \kappa)}$ with the effective mass being replaced by $m^\star=|\mu|/V^2$. Here $k_F=|\mu|/(\hbar V)$ is the Fermi wavevector and $n_s^0$ is the equilibrium surface density of charge carriers. At $q\gtrsim k_F$ the curve $\omega(q)$ enters the region of the inter-band damping and asymptotically tends to the line $\omega=Vq$ \cite{shu86,Vafek06,Wunsch06,Hwang07,Polini08,hil09}.

\begin{figure}
\begin{minipage}{7cm}
\begin{tabular}{c}
\includegraphics[width=5cm]{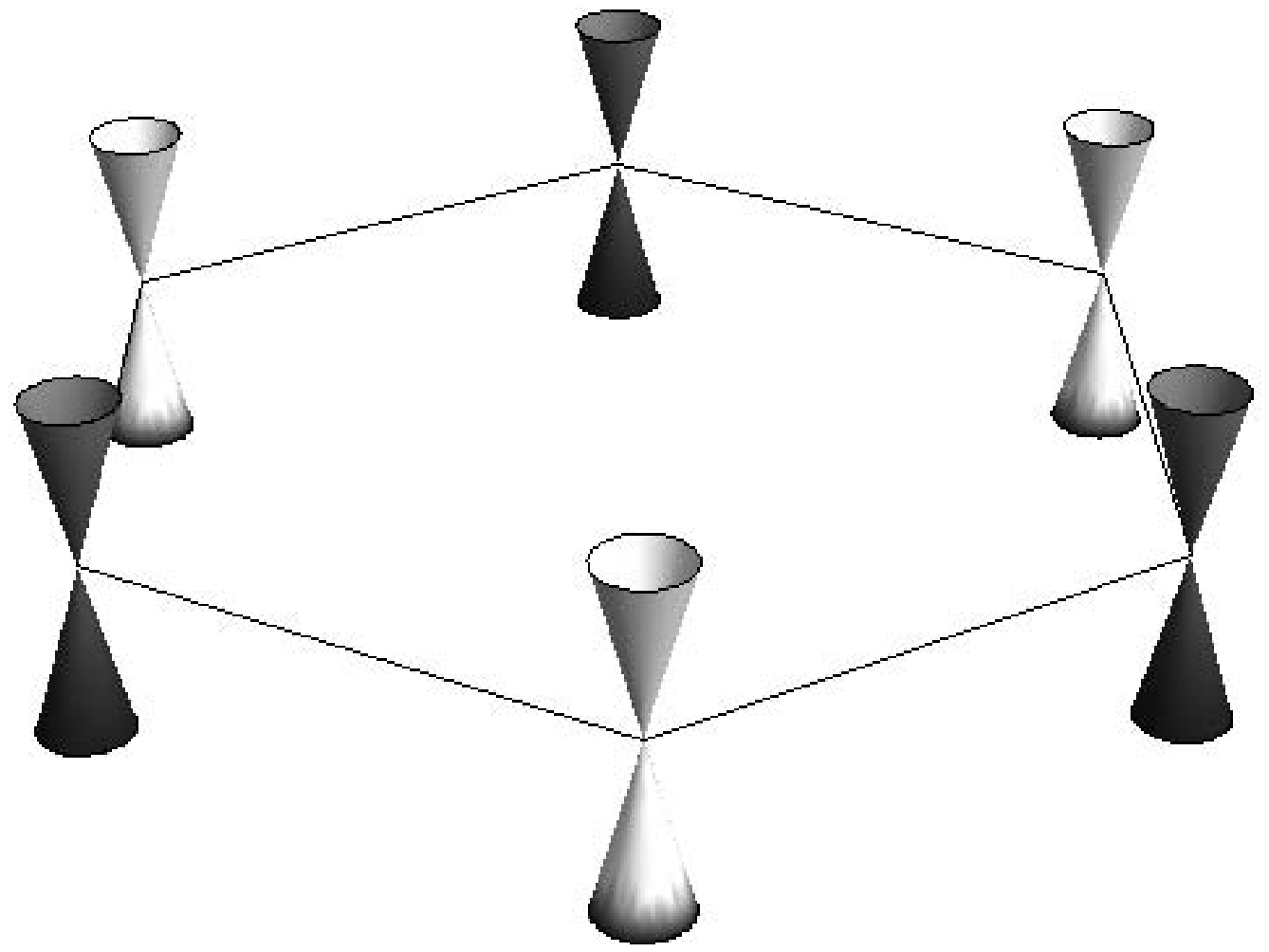}\\
(a)
\end{tabular}
\end{minipage}
\begin{minipage}{7cm}
\begin{tabular}{c}
\includegraphics[width=5cm]{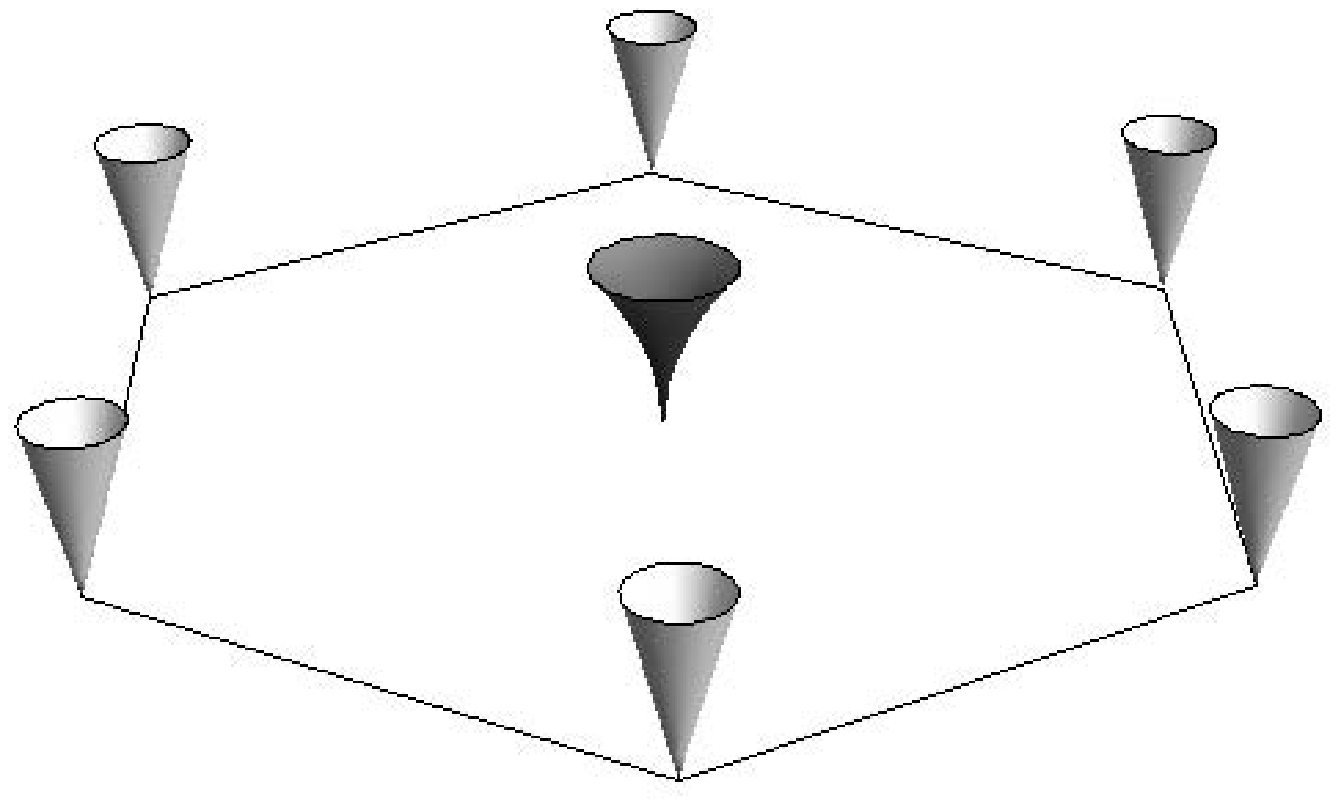}\\
(b)
\end{tabular}
\end{minipage}
\caption{\label{flowers} (a) The Dirac cones in the electron BZ and (b) the low-frequency 2D plasmon modes in the plasmon BZ.}
\end{figure}

The results outlined above are based on Eqs. (\ref{spectrum}), (\ref{epsilon00}) and are valid in the ``long-wavelength'' limit, when both the plasmon wavevector $q$ and the Fermi wavevector $k_F$ are small as compared to the reciprocal lattice vector $G\sim 1/a$. Here we study the 2D plasmon spectrum in graphene at the wavevectors ${\bf q}$ close to the corners of the BZ. The 2D plasmons propagate in the same periodic lattice as the 2D electrons and their spectrum $\omega_p({\bf q})$ should be a periodic function of ${\bf q}$ with the same hexagon-shaped BZ. Near the corners of the plasmon BZ ${\bf q=K}_i$ one can expect new low-frequency plasmon modes. Indeed, each 2D plasmon wavevector ${\bf q}\approx {\bf K}_i$ corresponds to an intervalley transition in the electron BZ ${\bf K}_{j}\to {\bf K}_{j'}$, for example, ${\bf q}= {\bf K}_1$ corresponds to the transition ${\bf K}_5\to {\bf K}_6$, ${\bf q}= {\bf K}_2$ -- to the transition ${\bf K}_6\to {\bf K}_1$, and so on. At ${\bf q}\approx {\bf K}_i$ the energy difference $E_{{\bf k}l}-E_{{\bf k}'l}$ in the denominator in Eq.~(\ref{epsilon00}) is close to zero, which leads to the new, {\it intra}-band {\it inter}-valley plasmon modes. In this Letter we show that these low-frequency plasmon modes have the linear dispersion,
\be
\omega_p(\textbf{q})=V_p |{\bf q}-{\bf K}_i|, \ \ |{\bf q}-{\bf K}_i|a\ll 1,
\label{intervalpl}
\ee
with the group velocity $V_p>V$. Figure \ref{flowers}b schematically shows the low-frequency plasmon mode in the BZ: the central dark-grey square-root ``flower'' and the light-grey ``flowers'' at the corners of the BZ correspond to the conventional \textit{intra}-valley 2D plasmon (\ref{longwavepl}) and  the \textit{inter}-valley plasmons (\ref{intervalpl}) respectively. For comparison, the Dirac cones at the corners of the electronic BZ are shown in Fig.~\ref{flowers}a.

In order to adequately calculate the graphene response at large wave-vectors $q\sim1/a$
one should go beyond equation (\ref{epsilon00}) and take into account the local field effects \cite{Adler62}. The electromagnetic response is then described by the matrix dielectric function
\[
\epsilon_{{\bf GG'}}({\bf q},\omega)=\delta_{{\bf GG'}}-\frac{2\pi e^2}{\kappa|{\textbf{q}+\textbf{G}}|}\chi_{{\bf G}{\bf G'}}({\bf q},\omega),\label{eq::eps}\\
\chi_{\bf GG'}(\textbf{q},\omega)=
\frac{g_s}{S}\sum_{\textbf{k}ll'}
\frac{f(E_{\textbf{k}l})-f(E_{\textbf{k}+\textbf{q},l'})}{E_{\textbf{k}l}-E_{\textbf{k}+\textbf{q},l'}+\hbar\omega+i0}
\langle\textbf{k}+\textbf{q},l'|e^{i(\textbf{q}+\textbf{G}')\textbf{r}}|\textbf{k}l\rangle
\langle\textbf{k}l|e^{-i(\textbf{q}+\textbf{G})\textbf{r}}|\textbf{k}+\textbf{q},l'\rangle,
\label{eq::chi}
\]
where ${\bf G},\,{\bf G}'$ are reciprocal lattice vectors and $\chi_{{\bf GG}'}$ is the polarizability tensor. The summation over $\textbf{k}$ in (\ref{eq::eps}) is performed over the whole BZ, $|\textbf{k}l\rangle$ are the Bloch functions and $E_{\textbf{k}l}$ is the corresponding energy dispersion. The 2D plasmon spectrum is determined by zeros of the determinant of $\epsilon_{{\bf GG'}}({\bf q},\omega)$,
\be
\det\|\epsilon_{{\bf GG'}}({\bf q},\omega)\|=0.
\label{eq::eps0}
\ee

From now on we use the tight binding approximation \cite{sai98} for the energy and the wavefunctions. Then the energy reads
$E_{{\bf k}l}=l\Delta |{\cal S}_{\bf k}|$, where $\Delta=2\hbar V/(\sqrt{3}a)$ is the full width of one band and
\be
\mathcal{S}_\textbf{k}=
1+2e^{i\sqrt{3}k_y a/2}\cos(k_x a/2).
\ee
The Bloch functions are
\be
|\textbf{k}l\rangle=\frac{1}{\sqrt{2N}}\sum_{\bf a}e^{i{\bf k a}}
\Bigl(\zeta^*_\textbf{k}\psi(\textbf{r}-\textbf{a},z)+l\psi(\textbf{r}-\textbf{a}-\textbf{b},z)\Bigr),
\label{eq::tbf}
\ee
where $N$ is the number of elementary cells inside the area $S$, $\zeta_{\bf k}={\cal S}_{\bf k}/|{\cal S}_{\bf k}|$, $\psi$ is the normalized atomic wavefunction and $z$ is the perpendicular coordinate. Using the wavefunctions (\ref{eq::tbf}) one can calculate the matrix elements in (\ref{eq::chi}),
\be
\langle {\bf k}l| e^{-i({\bf q}+{\bf G}){\bf r}} |{\bf k+q},l'\rangle=
\frac 12 M(|\textbf{q}+\textbf{G}|)
\left[
\zeta_{\bf k}\zeta^\star_{{\bf k+q}}+ll'
 e^{-i{(\textbf{q}+\textbf{G})}\textbf{b}}\right] ,\label{ME1}
\ee
where we assume that $\psi$ depends only on $|\textbf{r}|$ and $M(q)=\int d^3 r\ |\psi(\textbf{r},z)|^2 e^{i{\bf q r}}$. The integration in this formula is performed over the whole 3D space.

If $q\ll 1/a$, the terms $\chi_{{\bf GG}'}/|{\bf q}+{\bf G}|\sim M({\bf
q}+{\bf G})M({\bf q}+{\bf G}')/|{\bf q}+{\bf G}|$ in (\ref{eq::eps}) are small for all ${\bf G}$ and
${\bf G'}$ except  ${\bf G=G'=0}$. Then the general 2D plasmon dispersion
equation (\ref{eq::eps0}) is reduced to the one used in \cite{Vafek06,Wunsch06,Hwang07,Polini08,hil09}, $\det\|\epsilon_{\bf GG'}({\bf q},\omega)\|=\epsilon_{\bf 00}({\bf q},\omega)=\epsilon_{Dir}({\bf q},\omega)=0$.
If ${\bf q}$ is close to one of the vectors ${\bf K}_i$,
nine terms in the matrix $\chi_{\bf GG'}/|{\bf q+G}|$ 
give a noticeable contribution to the determinant of the
matrix $\epsilon_{\bf GG'}({\bf q},\omega)$. For example, if ${\bf q\simeq
K}_5$, the corresponding reciprocal lattice vectors are ${\bf G=0}$, ${\bf
G}_1$ and ${\bf G}_2$; for them $|{\bf q+G}|\simeq|{\bf K}_5|=K$. Thus for
${\bf q}$ close ${\bf K}_i$ the determinant of the infinite matrix
$\epsilon_{\bf GG'}({\bf  q},\omega)$ is reduced to the determinant of a
$3\times 3$ matrix with ${\bf G},\,{\bf G}'=\{{\bf 0},{\bf G}_1,{\bf G}_2\}$.

In what follows we assume that $\textbf{q}=\textbf{K}_5+\tilde{\textbf{q}}$, $\tilde{q}\ll k_F\ll K$ and the
temperature $T=0$. Then  one can show that the  inter-band contribution
$(l\neq l')$ to the polarizability tensor (\ref{eq::chi}) is negligible and we can use
the linear (Dirac) approximation for the energy in the vicinity of both
cones. Keeping in (\ref{eq::chi}) only the intra-band terms we get the following
expression for the polarizability tensor:
\[
\chi_{\textbf{GG}'}(\textbf{q},\omega)=
\frac{g_s}{4S}M^2(K)
\sum_{\textbf{k}}
\frac{\Theta_{\textbf{k}}-\Theta_{\textbf{k}+\textbf{q}}}
{E_{\textbf{k}+}-E_{\textbf{k}+\textbf{q},+}+\sigma\hbar\omega}
\eta_\textbf{G}(\textbf{k},\textbf{q})\eta_{\textbf{G}'}^*(\textbf{k},\textbf{q}),
\label{eq:eps3}
\]
where $\sigma=\sgn(\mu)$, $\Theta_\textbf{k}=\Theta(|\mu|-E_{\textbf{k}+})$, $\Theta$ is the Heaviside step function, $\eta_\textbf{G}(\textbf{k},\textbf{q})=\zeta^*_{\textbf{k}+\textbf{q}}\zeta_{\textbf{k}}
+e^{-i(\textbf{q}+\textbf{G})\textbf{b}}$ and ${\bf G},\,{\bf G}'=\{{\bf 0},{\bf G}_1,{\bf G}_2\}$; the chemical potential $\mu$ can be both positive and negative, i.e. our results are valid for both electron and hole gases.
Then introducing the notations $\textbf{k}=\textbf{K}_1+\tilde{\textbf{k}}$, the angle $\theta_1$ between the vectors $\tilde{\textbf{q}}$ and $\tilde{\textbf{k}}$ and the angle $\theta_2$ between the vector $\tilde{\textbf{q}}$ and $\textbf{K}_2=-\textbf{K}_5$, we get
\[
\chi_{\textbf{GG}'}
=
\frac{3g_s}{4\pi^2\hbar V}M^2(K)
\int_0^{k_F} \tilde{k}\,d\tilde{k}
\int_0^{2\pi}d\theta_1
\frac{|\tilde{\textbf{k}}+\tilde{\textbf{q}}|-\tilde{k}}
{-(|\tilde{\textbf{k}}+\tilde{\textbf{q}}|-\tilde{k})^2+(\omega/V)^2}
f_\textbf{G}(\theta_1+\theta_2) f^*_{\textbf{G}'}(\theta_1+\theta_2),
\label{eq:eps6}
\]
where $f_\textbf{G}(\theta)=(-e^{-2i\theta}+e^{-i\textbf{G}\textbf{b}})/\sqrt{6}$ for ${\bf G}=\{{\bf 0},{\bf G}_1,{\bf G}_2\}$ and
$\langle f|f\rangle=\sum_{\textbf{G}=\textbf{0},\textbf{G}_1,\textbf{G}_2}|f_\textbf{G}(\theta)|^2=1$. 

If $|\omega|<V\tilde q$, the integrand in (\ref{eq:eps6}) has poles on the $\theta_1$ axis and the functions $\chi_{\bf GG'}$ turn out to be complex. 
Similar to the standard 2D plasmons \cite{Vafek06,Wunsch06,Hwang07,Polini08,hil09} this corresponds to the single-particle intra-band absorption. At larger frequencies, $|\omega|>V\tilde q$, the denominator in (\ref{eq:eps6}) does not vanish and the functions $\chi_{{\bf GG}'}$ are real. In this region one can therefore expect a weakly damped (at low temperatures) low-frequency plasmon mode. As it will be seen from the result below, the frequency of the new plasmon mode is close to $V\tilde q$. Evaluating the leading term of the asymptotics of $\chi_{{\bf GG}'}$ with respect to the small parameter $(|\omega|/V\tilde q -1)\ll 1$ we get
\[
\epsilon_{\textbf{GG}'}
\simeq \delta_{\textbf{GG}'}-\frac{\beta}{\sqrt{|\omega|/V\tilde{q}-1}}
 f_\textbf{G}(\theta_2) f^*_{\textbf{G}'}(\theta_2),
\label{eq:eps8}
\]
where
\be
\beta=\frac{3g_s}{2\sqrt{2}}\frac{e^2}{\hbar V}\frac{k_F}{K}M^2(K).
\label{eq::beta}
\ee
Noticing that the matrix $\epsilon$ can be written as
$\epsilon=\mathbbm{1}-\textrm{const}\,|f\rangle\langle f|$, and using the formula $\det\left(\mathbbm{1}-\textrm{const}\,|f\rangle\langle f|\right)=1-\textrm{const}\,\langle f| f\rangle$ we finally get the spectrum of the inter-valley plasmon modes in the form (\ref{intervalpl}), where
\begin{equation}
V_p=V(1+\beta^2).
\label{eq:omega-inter-val}
\end{equation}
As it is seen from (\ref{eq::beta}) the factor $\beta$ is small as compared to unity.

Calculating the eigenvector corresponding to the inter-valley plasmon (\ref{intervalpl}), (\ref{eq:eps8}) we get the following expression for the potential of the plasmon mode
\be
\phi({\bf r},t)=\phi_0e^{i\tilde{\bf q}\cdot {\bf r}-i\omega t}\left[f_{\bf 0}(\theta_2)e^{i{\bf K}_5\cdot {\bf r}}+f_{{\bf G}_1}(\theta_2)e^{i{\bf K}_1\cdot {\bf r}}+f_{{\bf G}_2}(\theta_2)e^{i{\bf K}_3\cdot {\bf r}}\right];
\ee
the formula for the corresponding density fluctuation has a similar form.
One sees that the inter-valley plasmon with the wave vector ${\bf q}={\bf K}_5+\tilde{\bf q}$ is described by a plane wave with the (small) wavevector $\tilde{\bf q}$ and the amplitude periodically modulated with the wave vectors ${\bf K}_1$, ${\bf K}_3$ and ${\bf K}_5$.

In conclusion, we have found the new \textit{intra}-band \textit{inter}-valley low-frequency plasmon modes with the linear dispersion (\ref{intervalpl}) and the group velocity (\ref{eq::beta}). The appropriate description of these modes requires to take into account the local field effects.
The predicted modes do not exist in conventional 2D electron systems and are the unique feature of graphene. They could be observed using the Electron energy loss spectroscopy \cite{lan09}.

The work was supported by Deutsche Forschungsgemeinschaft and Swedish Research Council.

\end{document}